\documentclass[
    ,final            
  ]
  {aipproc}

\layoutstyle{6x9}

\begin{document}

\title{Probing Dark Energy in the Accelerating Universe with SNAP}

\author{Michael Schubnell (for the SNAP Collaboration)
\footnote{for a list of SNAP collaboration members see
  http://snap.lbl.gov}}
{address={Physics Department, University of Michigan, Ann Arbor, MI 48109}
}

\begin{abstract}
It has now been firmly established that the Universe is expanding
at an accelerated rate, driven by a presently unknown form of dark
energy that appears to dominate our Universe today.
A dedicated satellite mission has been designed to precisely map out
the cosmological expansion history of the Universe and thereby
determine the properties of the dark energy.
The SuperNova / Acceleration Probe (SNAP) will study thousands of
distant supernovae, each with unprecedented precision,
using a 2-meter aperture telescope with a
wide field, large-area optical-to-near-IR imager and high-throughput
spectrograph.
SNAP can not only determine the amount of dark energy with high
precision, but test the nature of the dark energy by examining how
its equation of state evolves.
The images produced by SNAP will have an unprecedented combination of
depth, solid-angle, angular resolution, and temporal sampling
and will provide a rich program of auxiliary science.

%
%
\end{abstract}

\maketitle


\section{Introduction}

Recent measurements of luminosity distance versus redshift of nearby
Type Ia supernovae by the Supernova Cosmology Project and the High-z
Supernova Team have determined that the expansion of the Universe is
accelerating \cite{Perlmutter, Riess}.
Furthermore, the results constrain the mass density, $\Omega_M$,
and the density of an unknown form of negative pressure energy,
$\Omega_\Lambda$, characterized by an equation of state $w\equiv p/\rho<-1/3$
causing the acceleration. This additional energy component, coined
dark energy, appears to dominate energy density and
dynamics of the Universe at the present epoch.

The evidence for dark energy is in remarkable concordance with other
observations. Measurements of small scale fluctuations in the
cosmic microwave background (CMB) radiation support
the supernova results and have determined that the Universe
is nearly flat \cite{Bennet}.
Observations of galaxy clustering \cite{Bahcall}
have shown that the fraction of the critical density consisting of
matter is $\Omega_M \approx$ 0.3, also
consistent with the results obtained from the supernova
measurements (Figure 1). Combined, these
results strongly suggest that - at the present epoch -  at least
70\% of the Universe's density is in the form of dark energy and
only approximately 30\% in some form of matter (which is mostly dark).

In its simplest form, dark energy might well be
Einstein's cosmological constant in the form of a vacuum energy
but numerous other theories have been proposed including
the possibility of slowly evolving scalar fields (so-called
quintessence models \cite{Ratra,Caldwell}.)
When combined with CMB and galaxy cluster measurements, a tight bound
on the dark energy equation of state $w$ can be extracted (assuming it is
constant over the expansion time, i.e. $dw/dz = 0$). Current constaints
on $w$ are consistent with dark energy being a cosmological constant
but also allow for a wide range of alternate models,
including those with a time dependent value of $w$ (Figure 1).
  
A precise measurement of dark energy properties requires
a much larger data set of supernovae than currently available
and a significant improvement of the systematic uncertainties
in the measurements over current experiments.
A definitive program to study dark energy with supernovae
must provide a high degree of statistical and systematic rigor \cite{Kim1}.
Furthermore, the greatest sensitivity to cosmological parameters is
obtained with measurements extending from the present epoch
 of acceleration into the matter dominated
deceleration phase \cite{Linder}. Because measurements
of the highly redshifted light from very distant
 supernovae require sensitivity into the near infrared (NIR), such a program 
can only be achieved in space, unhindered by absorption in the earth's
atmosphere.

\begin{figure}
  \includegraphics[height=.58\textwidth]{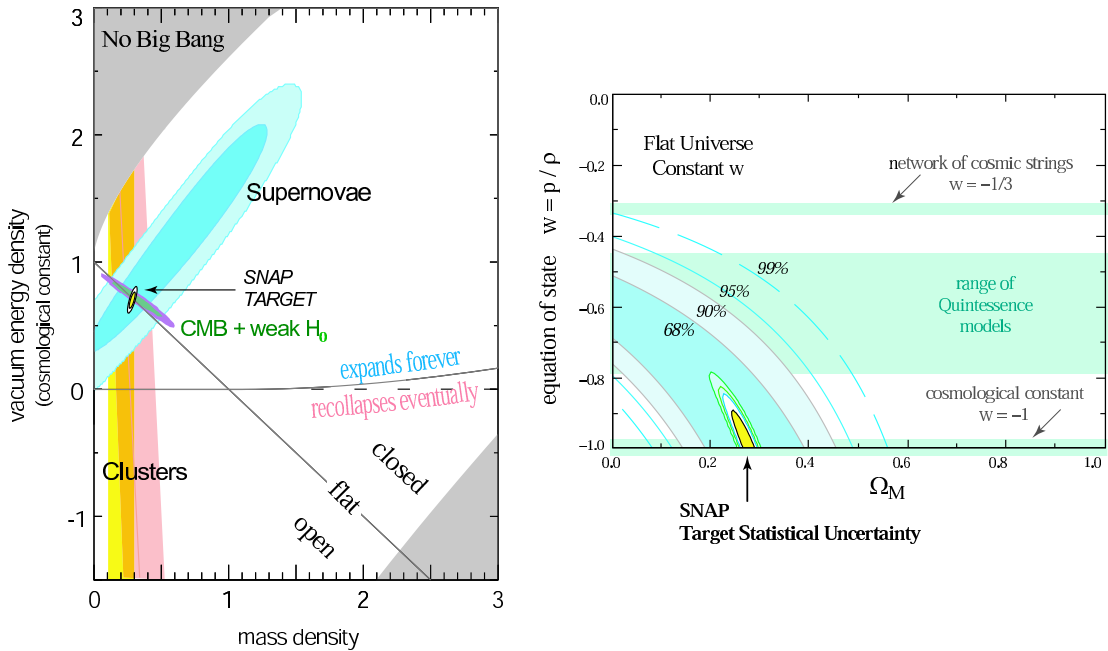}
  \caption{Left figure --
Confidence regions in the $\Omega_{\Lambda}-\Omega_{M}$ plane
for supernovae \cite{Perlmutter}, galaxy cluster \cite{Bahcall},
and CMB \cite{Bennet} data. The consistent overlap is compelling
evidence for a geometrically flat, dark energy dominated Universe.
Also shown (in both figures) is the expected confidence
region from the SNAP satellite for a flat $\Omega_M=0.28$ Universe.
Right figure --
Constraints on the equation-of-state parameter $w$ from the
Supernova Cosmology Project \cite{Perlmutter}. Shown are confidence
regions in the $\Omega_{\Lambda}-w$ plane for an energy
density component $\Omega_{\Lambda}$ characterized by $w=p/\rho$.
If the dark energy is Einstein's cosmological constant, then $w=-1$.
Also shown are $w$ predictions for other dark energy models.
}  
\end{figure}

\section{Satellite and Mission}

The primary goal of the SNAP mission is to measure
cosmological parameters with a precision that will allow
to distinguish between different dark energy models. 
For this supernova observations provide a proven and well
understood cosmological tool. The essential measurement for
this purpose is a comparison of luminosity distance to redshift
providing information on the scale size as a function of
expansion time. With a precisely calibrated data set of several
thousand Type Ia SNe with $z=0.1 - 1.7$ the
ex\-pansion history of the Universe can be reconstructed back
to more than 70\% of its age.

It has been shown that Type Ia supernovae have uniform peak
$B$-band brightness when their light curves are corrected for a stretch
factor which describes the relation between absolute brightness and
explosion \cite{Perlmutter}. However, to fully standardize the SN peak
brightness, a variety of additional observations must
be made. Color measurements throughout the light curve for instance
provide constraints on host-galaxy environment and galactic extinction
and spectra obtained near maximum brightness allow
identification of the explosion as a Type Ia through characteristic
features (e.g. SiII at 6150 \AA).

The SNAP satellite and mission design has been optimized for efficient
supernova detection and high quality follow-up measurements.
The combination of a three mirror 2-meter telescope  
and a $\approx$ 600 million pixel optical to near infrared imager with a
large 0.7 square degree field of view will allow 
discovery and follow up of many supernovae at once.
The imaging system comprises 36 large format (3.5k $\times$ 3.5k)
CCDs and the same number of 2k $\times$ 2k HgCdTe infrared sensors.
Both the CCDs and the NIR detectors are placed in four symmetric 3 $\times
$ 3 arrangements as shown in Figure 2. Both the imager and a low
resolution (R$\approx$100) high-throughput spectrograph cover the waveband
from 350 to 1700 nm, allowing detailed characterization of supernovae
out to $z=1.7$. This deep reach in redshift is essential to the mission
as it will allow to resolve degeneracies in cosmological parameters
and to discriminate between models of dark energy.

\begin{figure}
{
\begin{minipage}[t]{3.5in}
  \center{\includegraphics[height=2.in]{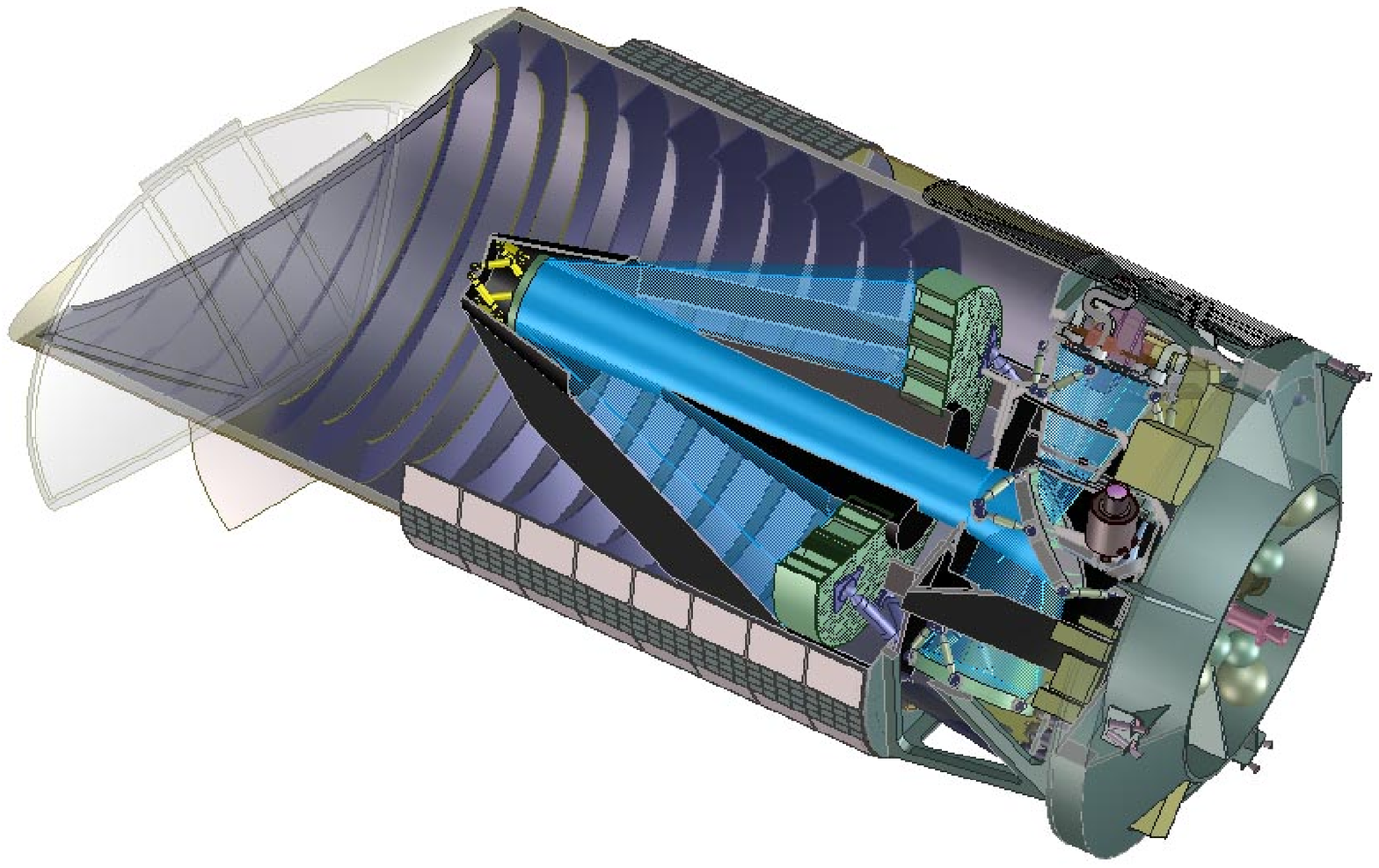}}
\end{minipage}
\begin{minipage}[t]{2.5in}
 \includegraphics[height=2.in]{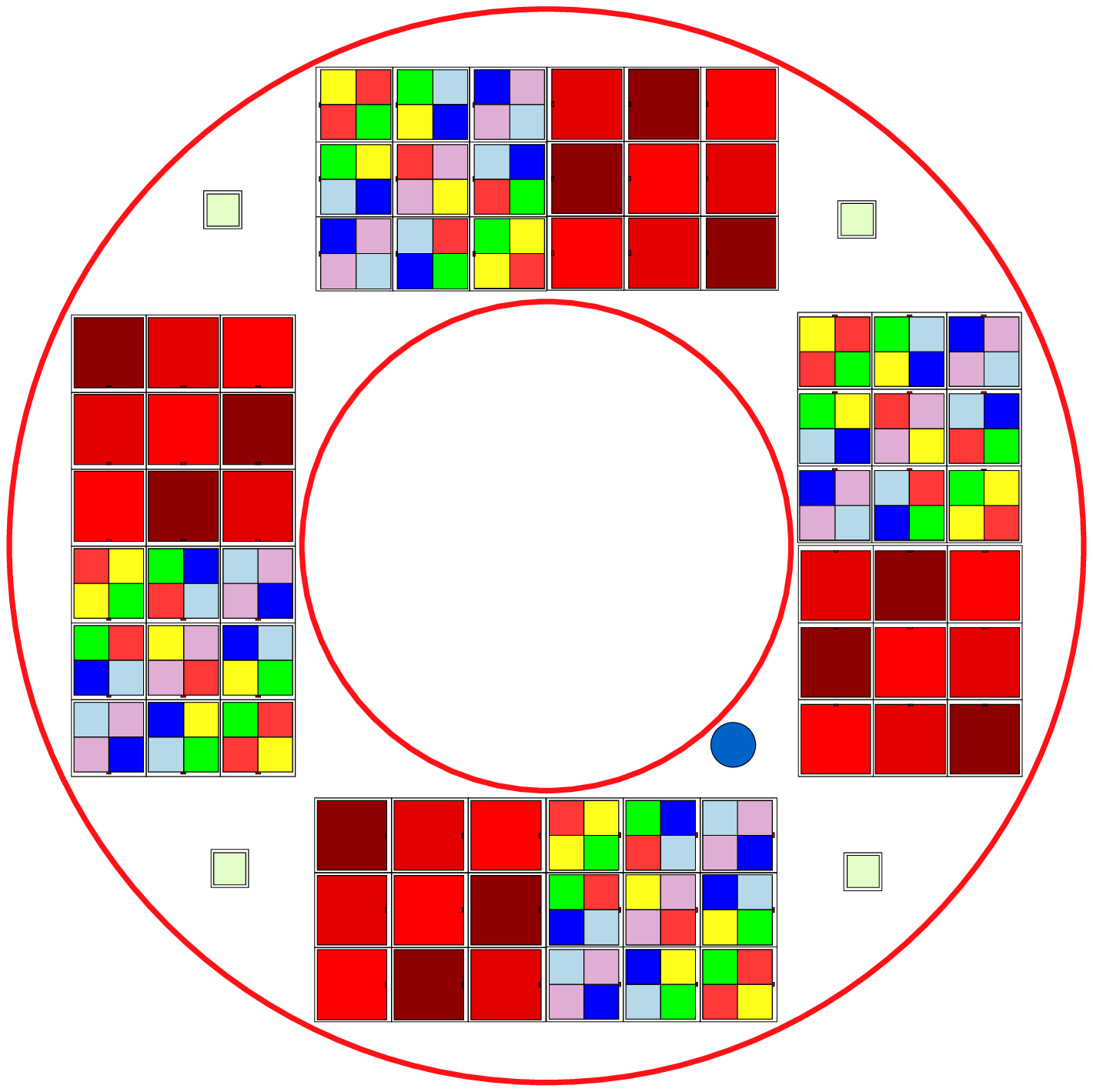}
\end{minipage}}
  \caption{Left figure -- Cross-sectional view of the SNAP
    satellite. Right figure -- The SNAP focal plane. For detailed
    description of satellite and anstrument see \cite{Lampton,
    Lampton2}
}
\end{figure}

Nine special filters fixed above the imaging sensors will provide overlapping
red-shifted $B$-band coverage from 350 -- 1700 nm. As SNAP repeatedly
steps across its target fields in the north and south ecliptic poles, every
supernova will be seen in every filter in both the visible and NIR.
Because of their larger linear size, each NIR filter will be visited
with twice the exposure time of the visible filters.  This, combined with the
time-dilated light curve, will ensure that Type Ia supernovae out to 
redshift 1.7 will be detected with a S/N $>$ 6 at least 2 magnitudes below
peak brightness \cite{Lampton,Lampton2}. 

\section{SNAP Science}

SNAP will conduct two
primary surveys, a $\sim15$ square degree ultra-deep ($m_{AB} \sim 30$
for point sources)
supernova survey, and a $\sim 300$ square degree deep 
($m_{AB} \sim 27.8$ for point sources) weak lensing survey.
With this wealth of detailed data, SNAP will construct a Hubble
diagram with unprecedented control over systematic uncertainties,
addressing all known and proposed sources of error.  The first goal is
to provide precision measurements of the cosmological parameters:  the
matter density, $\Omega_M$, will be measured to $\pm$0.02,
while $\Omega_\Lambda$, and the curvature parameter, $\Omega_k$,
will both be determined to an accuracy of $\pm$0.04.  The SNAP
measurements will be largely orthogonal to the CMB measurements
in the  $\Omega_M -\Omega_\Lambda$ plane, and the curvature
measurement at z $\approx$ 1 will test cosmological models
by comparison with the CMB determination at $z\approx1000$.
SNAP's science reach will then extend to an exploration of the nature
of  the dark energy, measuring the present equation of state, $w$, to 5\%.
Of even more interest is a determination of $w$ as a function of
redshift.  SNAP's tight control of systematics and high statistics
in each redshift bin allows determination of the dynamical variation
of $w´$.

To complement its supernova cosmology observations, SNAP will conduct a 
wide-area weak lensing survey. These weak lensing observations provide
important independent measurements and complementary determinations 
of the dark matter and dark energy content of the Universe. They will
substantially enhance SNAP's ability
to constrain the nature of dark energy\cite{Rhodes}. SNAP weak 
lensing observations benefit enormously from the high spatial
resolution, the accurate photometric redshifts, and the very high surface 
density of resolved galaxies available in these deep observations.

Although the SNAP mission is tailored for supernova and weak lensing
observations, with the large survey field, depth, spatial resolution,
temporal sampling and wavelength coverage into the infrared,
the resulting
data sets will provide a rich program of auxiliary science.
Here we highlight a few selected areas where the large area deep-field
observations are expected to significantly impact our
understanding of the Universe:

\begin{itemize}

\item
Galaxies -- Within the ultra-deep 15 square degree survey area, SNAP
will obtain accurate photometric redshift measurements for at least 
$5\times10^{7}$ galaxies to $z$=3.5. This will provide a
unique opportunity to study the evolution of galaxies through more
than 90\% of the age of the Universe.

\item
Galaxy clusters -- Galaxy clusters, the most massive bound objects
and probably largest structures in the Universe, provide important
probes of our understanding of structure formation. The SNAP surveys
will provide detailed information on roughly 15,000 galaxy clusters
with masses above $5\times10^{13}$ M$_\odot$.

\item
Quasars -- NIR photometry extends
the redshift range for quasar discovery using colors and dropout surveys.
SNAP will be able to detect quasars beyond redshift 10, and to probe
the quasar luminosity function to 100 times fainter than the
brightest quasars.

\item
GRBs -- Gamma-ray bursts continue to pose a great mystery. Recent
observations point to GRBs as the product of
core-collapse of super-massive short lived stars. If so, then
GRBs may trace the star formation rate and thus 
GRBs coincident with the epoch of first
stars formation are expected. The most distant GRB currently known
occurred at redshift of 3.4. SNAP
will be able to identify GRB afterglows to $z = 10$.

\item
Re-ionisation -- Most likely the re-ionisation of the Universe
did not occur as a single instant in time, but rather as a complex
process happening at slightly different epochs in
different parts of the Universe. By identifying many quasars
and galaxies to $z=10$, SNAP will map the epoch of re-ionization in
unprecedented detail.

\item
Gravitational lensing -- The high spatial resolution of
SNAP NIR observations will enable the discovery of a large number of
new strong lenses.
In weak lensing measurements, SNAP spatial resolution and
NIR sensitivity will allow the use of a huge number of faint,
high-redshift background galaxies. With these galaxies, it will be
possible to extend weak lensing studies to lower mass objects,
and to study lens objects beyond $z = 1$ -- measurements which
are impossible from the ground.

\end{itemize}

\section{Conclusion}

SNAP presents a unique opportunity to probe the
dark energy and advance our understanding of the Universe.
It will discover
and precisely measure thousands of supernovae of Type Ia and will
provide a combination of depth, solid angle and angular resolution
heretofore unachieved.  From the data collected,
it will be possible to precisely measure the
equation of state of the Universe, measure the history of its
accelerations and decelerations and to study the nature of
dark energy, which is causing the current acceleration of 
the expansion of the Universe.  
SNAP will be able to measure 
the equation of state, $w$, of the Universe as well as its
variation over time, $w'$.
This detailed knowledge will allow to distinguish
between different models for the nature of the dark energy
and lead to deeper understanding of the Universe.


\begin{theacknowledgments}
  This work was supported by the U.S. Department of Energy.
\end{theacknowledgments}


\bibliographystyle{aipprocl} 

\bibliography{schubnell_snap}

\IfFileExists{\jobname.bbl}{}
 {\typeout{}
  \typeout{******************************************}
  \typeout{** Please run "bibtex \jobname" to optain}
  \typeout{** the bibliography and then re-run LaTeX}
  \typeout{** twice to fix the references!}
  \typeout{******************************************}
  \typeout{}
 }

\end{document}